\documentclass[a4paper,11pt]{article}
\pdfoutput=1 
\usepackage{jinstpub} 
\usepackage{lineno}
\usepackage{url}
\usepackage{graphicx}
\usepackage{caption}
\usepackage{float}
\usepackage{subfig}
\usepackage{appendix}
\usepackage{hyperref}
\usepackage{xspace} 

\title{\boldmath Cryogenic RPWELL: a novel charge-readout element for dual-phase argon TPCs}

\author[a, \dagger]{A. Tesi\note{Corresponding author}}
\author[a,b, \dagger]{, S. Leardini}
\author[a]{, L. Moleri}
\author[b]{, M. Morales}
\author[b]{, D. Gonzalez-Diaz}
\author[a]{, A. Jash}
\author[a]{, A. Breskin}
\author[a]{and S. Bressler}

\affiliation[a]{Department of Particle Physics and Astrophysics, Weizmann Institute of Science, Rehovot, Israel}
\affiliation[b]{Instituto Galego de Física de Altas Enerxías, Univ. de Santiago de Compostela, Santiago de Compostela, Spain}
\affiliation[\dagger]{Contributed equally}

\emailAdd{andrea.tesi@weizmann.ac.il}

\abstract{The first operation of a cryogenic Resistive Plate WELL (RPWELL) detector in the saturated vapor of liquid argon is reported. The RPWELL detector was composed of a Thick Gas Electron Multiplier (THGEM) electrode coupled to a metallic anode via Fe$_2$O$_3$/YSZ ceramics (Fe$_2$O$_3$ in weight equal to 75$\%$), with tunable bulk resistivity in the range 10$^{9}$ - 10$^{12}$ $\Omega\cdot$cm. 
The detector was operated at liquid argon temperature in saturated argon vapor (90~K, 1.2~bar) and characterized in terms of its effective charge gain and stability against discharges. Maximum stable gain of G$\approx$17 was obtained, without discharges.
In addition, preliminary results from novel 3D-printed thermoplastic plates doped with carbon nanotubes are presented. 
}


\begin{document}
\maketitle
\flushbottom

\section{Introduction}
\label{sec:intro}
\begin{figure}[h]
    \centering
    \includegraphics[width=0.9\linewidth]{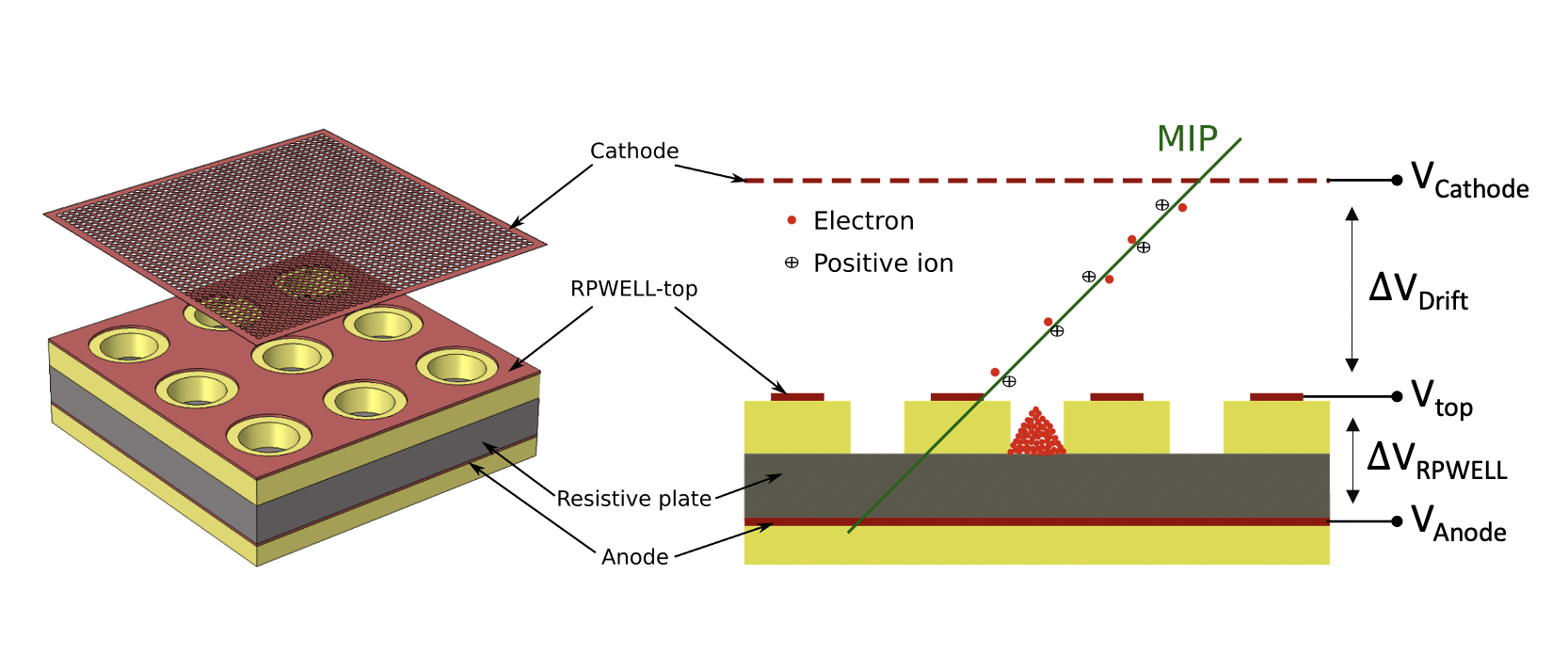}
    \centering
    \caption{Schematic drawing of an RPWELL detector. In the present work, the drift gap was equal to 15~mm and the single-sided THGEM electrode was 0.8~mm thick with 0.5 mm diameter holes distributed in a hexagonal pattern with 1 mm pitch, 0.1 mm hole-rim.}
    \label{fig:RPWELL_scheme}
\end{figure}

In recent years, the Thick Gaseous Electron Multiplier (THGEM) has become a well-established gaseous radiation detector \cite{Chechik:2004wq, Breskin:2008cb, Bressler_2023}. The possibility to use it for amplifying ionization charges in a gas medium at different pressures and temperatures, its robustness, and its low cost turned this detector into a favorable option for several experiments requiring large area coverage at modest spatial and energy resolutions. Some examples are: Ring-Imaging Cherenkov (RICH) detectors \cite{Chechik_2005, Alexeev:2010nyj, Breskin:2010wi}, digital hadronic calorimetry (DHCAL) \cite{Arazi_2012, Bressler_2013}, rare events searches \cite{Breskin_LHM,Arazi:2015prw,Bondar:2011he, Bondar_2013} and civil \cite{Israelashvili:2014ona, Israel_2}  and medical applications \cite{Duval:2009hi,AZEVEDO2013551}. For a recent review on THGEM detectors see \cite{Bressler_2023}. Amongst the THGEM derivatives, the Resistive-Plate WELL (RPWELL) \cite{Rubin:2013jna} was studied at room temperature, mostly for sampling elements in Digital Hadronic Calorimetry (DHCAL)\cite{Bressler_2019,Renous_2020, Renous_2022}, due to its high gain and protection against discharges (or: discharge-quenching properties).
The RPWELL detector shown schematically in Fig.~\ref{fig:RPWELL_scheme} consists of a single-sided THGEM electrode coupled to an anode via a  plate of high bulk resistivity. It outperforms both its discharge-unprotected counterparts (e.g. THGEM, THWELL \cite{Bressler:2013cxa}), and the discharge-protected Resistive WELL (RWELL) \cite{Arazi:2013hdn}, which includes a thin coating of resistive material applied onto an insulating layer. Charges accumulating on the resistive plate result in a local and temporary decrease of the electric field intensity, thus potentially, quench the discharge energy \cite{Jash:2022bxy}. Resistive plates with a bulk resistivity value in the range 10$^9$
~<~R$_\mathrm{V}$~<~10$^{12}~\Omega\cdot$cm \cite{Gonzalez-Diaz:2012vwx, Santonico:1981sc} are typically used to obtain sufficient quenching, yet preserving high gain under high incoming radiation rate flux.

\noindent
The properties of the RPWELL detector were assessed also at liquid xenon (LXe) temperature, 163~K, in gaseous Ne:5$\%$CH$_4$ \cite{Roy_2019}. The resistive plate was made of a ferrite ceramic \cite{Morales:2013rka} of bulk resistivity R$_\mathrm{V}\approx$10$^{11}~\Omega\cdot$cm at 163~K, showing discharge quenching capabilities also at this cryogenic conditions. To explore the possibility of operating an RPWELL at liquid argon temperature, 90~K, a major challenge was to find or engineer appropriate materials having the aforementioned bulk resistivity. It is known that most resistive materials undergo an exponential increase of their bulk resistivity at low temperatures (Arrhenius' law \cite{Gonzalez-Diaz:2005vxe}), usually becoming unsuitable for operation in cryogenic particle detectors. This motivated the development of new materials, in collaboration with the Instituto Galego de Fisica de Altas Enerxias in Santiago de Compostela. A novel ceramic material based on Yttria-Stabilized Zirconia (YSZ) and iron oxide (Fe$_2$O$_3$) was engineered for operation at 90~K, as reported in \cite{Tesi_2023, Olano}. Compared to \cite{Roy_2019}, the ceramics production, in particular the post-processing stage, has been improved, and the concentration range extended to cover all temperatures of potential interest down to liquid argon (LAr) temperature.\\
\noindent

\section{Experimental Setup and Methodology}
\label{sec:setup}
Experiments were conducted in a dedicated cryostat, WISArD, described in detail in \cite{Tesi:2021kan, Erdal_phd, Tesi_2023_rwell}. The RPWELL assembly consisted of a 3x3~cm$^2$ area and 0.8~mm thick single-sided THGEM (0.5~mm diameter holes distributed in a hexagonal pattern with 1~mm pitch, 0.1~mm hole-rim) pressed onto a resistive plate. The latter is attached to a plain metallic anode using cryogenic conductive epoxy\footnote{Master Bond EP21TDCS-LO}. Charge signals are read out from the anode. 

\paragraph{Ceramic plate}
A sample of 2~mm thick YSZ ceramic plates containing 75$\%$ Fe$_2$O$_3$ was used. The bulk resistivity R$_\mathrm{V}$ at 90 K was 8$\times$10$^{10}$~$\Omega\cdot$cm at $\sim$20 V (low-voltage drop, amenable to present conditions). 
A detailed account of the properties of these ceramic materials at cryogenic temperatures can be found in \cite{Olano}. \\

\paragraph{3D-printed resistive materials}
3D-printed thermoplastics enriched with carbon nanotubes (CNT) can also be produced at the desired value of conductivity. In collaboration with the 3D functional and printing center of the Hebrew University of Jerusalem, several samples of resistive plates were fabricated for testing at 90~K. A sample of 1.6~mm thick Acrylonitrile Butadiene Styren (ABS)\footnote{\href{https://www.3dxtech.com/product/3dxstat-esd-abs/?attribute_pa_diameter=2-85mm&attribute_pa_weight=750g&attribute_pa_color=black}{3DXSTAT ESD-ABS}} was investigated and its bulk resistivity R$_\mathrm{V}$ was found to be 3$\times$10$^{12}$~$\Omega\cdot$cm at 298 K and 7$\times$10$^{12}$~$\Omega\cdot$cm at 77 K.\\ 

\noindent
For the operation of the RPWELL, it is possible to estimate the maximal voltage drop $\Delta V$ produced by an event under the assumption that all the primary charge goes into a single hole. 
The avalanche charge generated by a 4~MeV-$\alpha$ in gaseous argon is Q = n$_p$$\times q\times$G = 0.24~pC, with n$_p$ $\approx$10$^5$ primary electrons, q = 1.6$\times$10$^{-19}$C and detector gain G = 15. Given the hole diameter d, the hole area reads A = $\pi\times(\frac{d}{2})^2 \approx$0.2~mm$^2$.
Under parallel-plate capacitance approximation (C = $\frac{\epsilon_0\epsilon_{Ar}A}{t}$ = 2.21 fF, $\epsilon_0\epsilon_{Ar}$ = 8.859$\times$10$^{-12}$ $\frac{C}{Vm}$ and electrode thickness t = 0.8 mm), $\Delta V = \frac{Q}{C} \approx$108 V. The corresponding uniform electric field is E $\approx$135 V/mm. 

\begin{figure}[h]
 \centering
    \includegraphics[width=12cm]{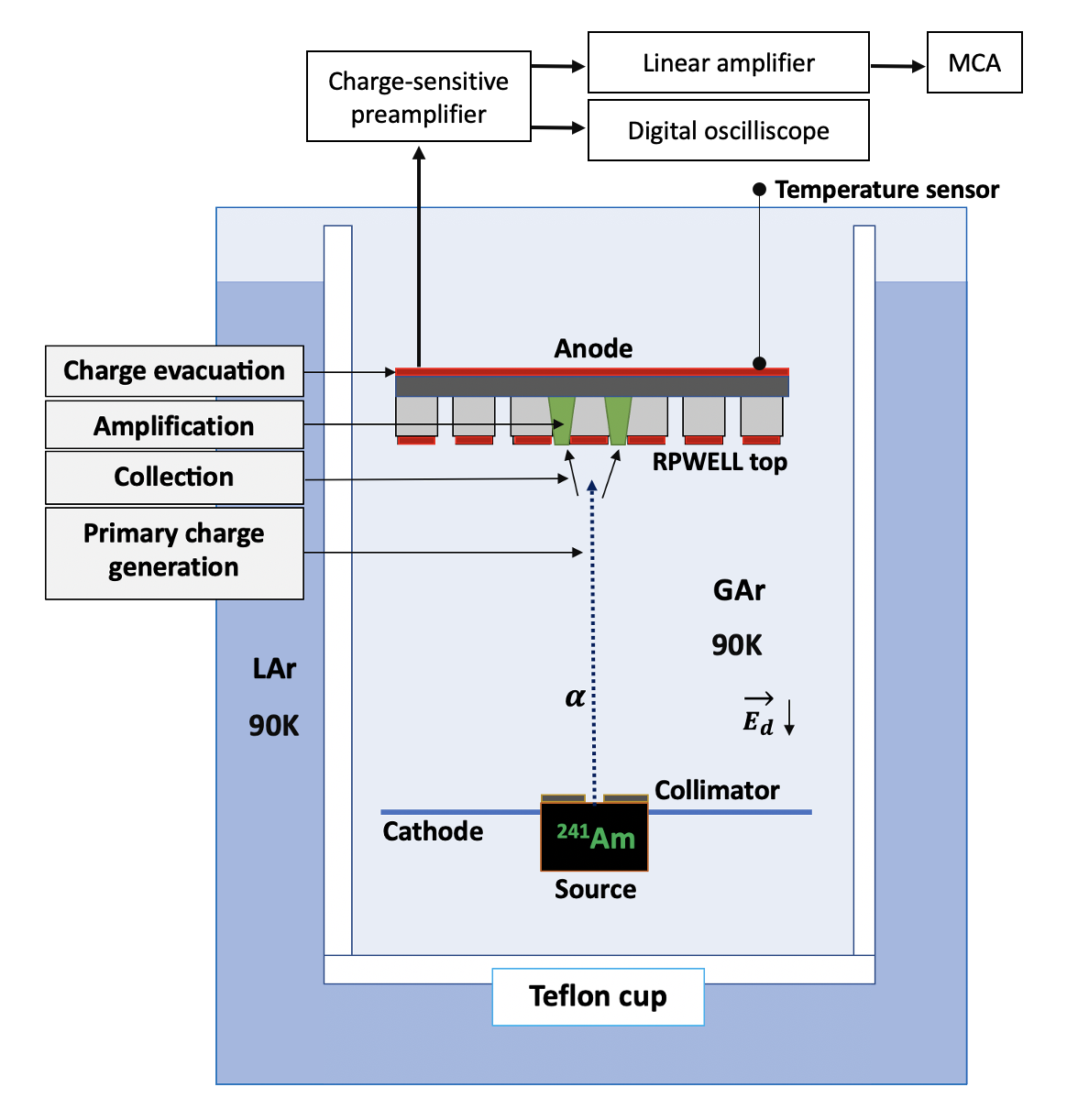}
    \caption{Scheme of the cryogenic experimental setup for investigating the RPWELL detector. The assembly was composed of a metallic-plate cathode equipped with a collimated $^{241}$Am source fixed at 15 mm from the amplification stage (corresponding to E$\mathrm{_d}$ = 0.5 kV/cm). The multiplier consists of a 0.8 mm thick single-sided THGEM coupled to an anode via a resistive plate. The assembly was inserted into a Teflon cup and operated in the liquid argon vapor at 90~K, 1.2~bar. The involved physical processes are listed.}
    \label{fig:RPWELL_setup}
 \end{figure}

\paragraph{Setup}
A drift field of E$_\mathrm{d}$~=~0.5~kV/cm was implemented in the region between the RPWELL and the metallic cathode (Fig.~\ref{fig:RPWELL_scheme}) located 15 mm away from the top THGEM electrode. The detector was irradiated with a $^{241}$Am source, at a rate of about 10~Hz. The source was installed on the cathode plate; it was collimated with a 4~mm thick collimator of 5~mm diameter aperture, attenuated (using 12~$\mathrm{\mu}$m thick mylar foil) down to $\sim$4~MeV - as to keep the alpha-particle range at $\approx$10~mm.
Alpha-induced ionization electrons were collected from the drift region along the field lines into the RPWELL holes, where charge multiplication occurred. Avalanche electrons are evacuated to the anode through the resistive plate, while ions back-flow to the RPWELL top electrode. A signal is formed on the anode by the movement of electrons and ions, in accordance with Shockley-Ramo theorem \cite{Bhattacharya:2018sqx}. The assembly was fixed with a Teflon cup, and liquid argon was filled into the cryostat up to the cup's edges. In such a way, thermal equilibrium was granted between the liquid phase outside the cup and the saturated vapor inside. During operation, the gas was recirculated and purified with a hot getter\footnote{Entegris HotGetter PS3-MT3-R-2}, at a flow of $\approx$2~l/h, to grant a nominal impurity content of <~1ppb (for more details see the methodology section in \cite{Tesi_2023_rwell}). The temperature in the proximity of the amplification element was constantly monitored with a temperature sensor installed on the back side of the readout anode. The experimental setup is depicted in Figure~\ref{fig:RPWELL_setup}.
 
\noindent
The cathode and the RPWELL top were biased by a high voltage power supply\footnote{CAEN N1471H}, and the anode was grounded via a charge-sensitive preamplifier\footnote{Cremat: Model CR-110 with CR-150-R5 evaluation board}. The preamplified signals were processed with a linear amplifier\footnote{Ortec Model 450} and digitized by a multi-channel analyzer\footnote{Amptek MCA 8000D}. Signals were also acquired with a digital oscilloscope\footnote{Tektronix MSO 5204B} and analyzed offline with dedicated Matlab scripts. Power-supply currents from the electrodes were recorded by a digitizer\footnote{NI USB-6008} and monitored using LabVIEW SignalExpress \cite{Labview}.

\paragraph{Methodology}
The measurement methodology follows closely the one described in \cite{Tesi_2023_rwell} and consists of the following steps:
\begin{itemize}
    \item Purification and thermal stabilization of the system ({12 hours})
    \item Power adjustment and gain stabilization ({minimum 5 hours}) 
    \item Measurement of the charge distribution in charge-collection mode and of the collection peak position from the RPWELL top electrode
    \item Effective gain estimation - charge-distribution peak position measured from the anode normalized to the collection peak. 
    \item Discharge monitoring through the power supply currents (Labview SignalExpress).
\end{itemize}
For the gain stabilization measurement, the linear amplifier\footnote{Ortec Model 450} was set to a shaping time of 10~${\mathrm{\mu}}$s. Histograms were recorded using the MCA, during 120~s. All the spectra were normalized to their area and to the maximum at $\mathrm{\Delta V_{RPWELL}}$ = 2.7~kV.
A data set of 5k waveforms was recorded for each measurement with the oscilloscope. 

\section{Results}
\label{sec:results}

A systematic study was carried out with the 75$\%$ Fe$_2$O$_3$-RPWELL. 
Preliminary results measured with an ABS-RPWELL are presented as well.

\subsection{Gain stabilization}
\label{subsec:gain_stabilization}

\begin{figure}[h]
    \includegraphics[height=11cm, width=16cm]{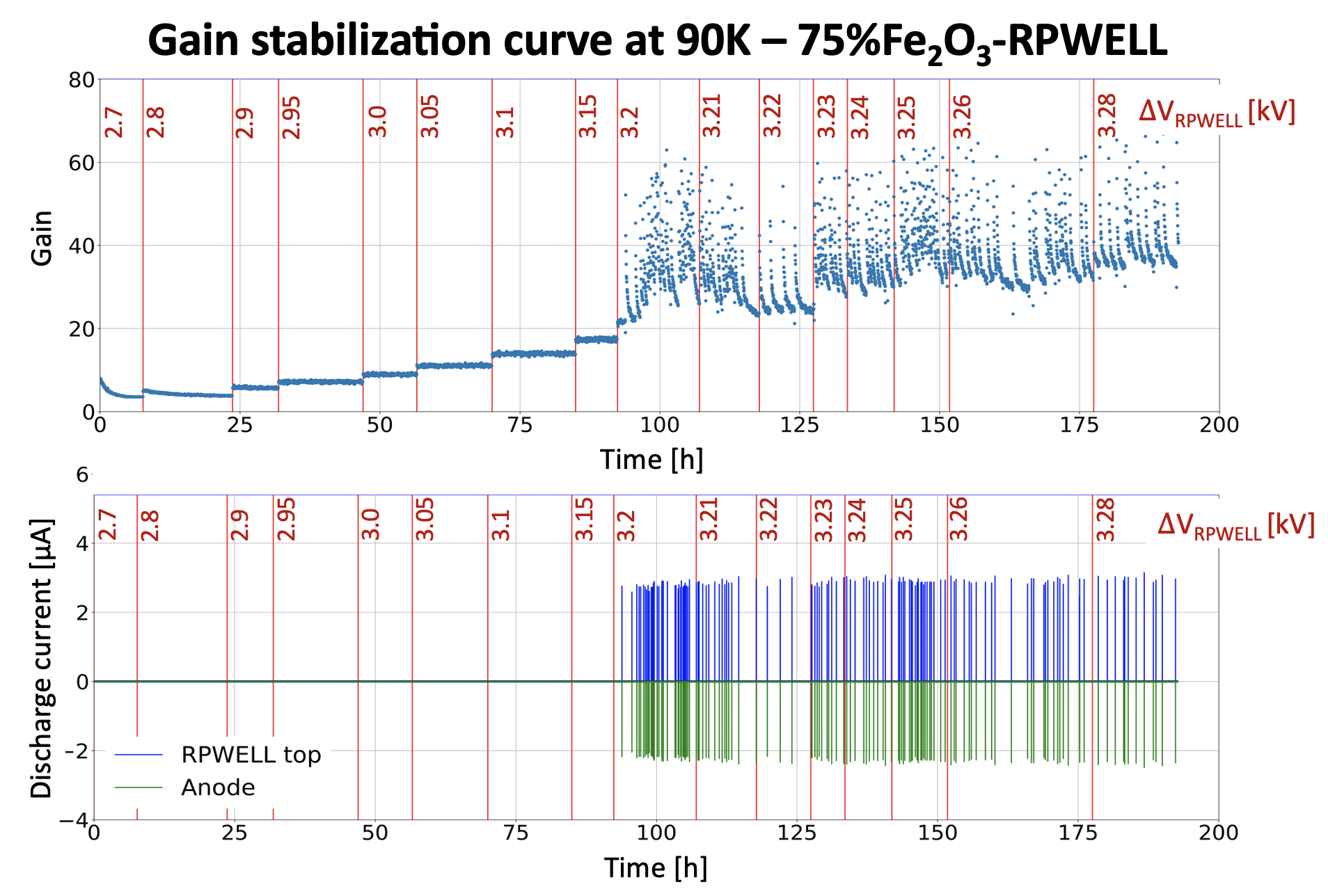}
    \caption{{\bf{Top}}: Gain stabilization curve over approximately 200 hours of a 75$\%$ Fe$_2$O$_3$-RPWELL operated at 90~K, 1.2~bar. The voltage configurations are indicated by red vertical bars with the corresponding values of $\Delta V_\mathrm{RPWELL}$. {\bf{Bottom}}: Power supply currents recorded from the RPWELL top electrode (blue) and anode (green) for the same measurement. Note the presence of discharges above $\Delta$V$\mathrm{_{RPWELL}}$~=~3.2~kV.}
    \label{fig:RPWELL_stabilization}
 \end{figure}

\noindent
The drift field was set at E$\mathrm{_d}$~=~0.5~kV/cm and the voltage across the 75$\%$ Fe$_2$O$_3$-RPWELL; $\Delta$V$_{\mathrm{RPWELL}}$ was varied between 2.7~kV to 3.28 kV. The gain evolution during the scan is reported in Fig.~\ref{fig:RPWELL_stabilization}, top. To correlate gain instabilities with discharges occurring in the RPWELL, the power supply currents monitored from the cathode, RPWELL top, and anode, are reported in Fig.~\ref{fig:RPWELL_stabilization}, bottom. 

 \noindent

 As can be seen, a gain drop of the order of 60\% was recorded during the initial stabilization at $\Delta$V$\mathrm{_{RPWELL}}~=~$2.7~kV. This is attributed to the charging-up of the dielectric substrate of the THGEM \cite{Renous_2017, Pitt_2018}. Stable operation was recorded in the range 2.7~kV < $\Delta$V$\mathrm{_{RPWELL}}$~<~3.2~kV. 
 Unstable operation, as indicated by correlated spikes between the gain measurements and the current ones, were  
 observed when the detector was operated in the range 3.2~kV < $\Delta$V$\mathrm{_{RPWELL}}$~<~3.28~kV. Nevertheless, the detector was able to sustain these discharges. The measurement was stopped at $\Delta$V$\mathrm{_{RPWELL}}$~=~3.28~kV, above which discharges started piling-up causing a DC voltage drop across the amplification structure (detector tripping). Similar behavior was recorded with the ABS-RPWELL.

\subsection{Typical spectra}
\label{subsec:typical_waveforms}
\begin{figure}[h]
\center
    \includegraphics[width=8cm]{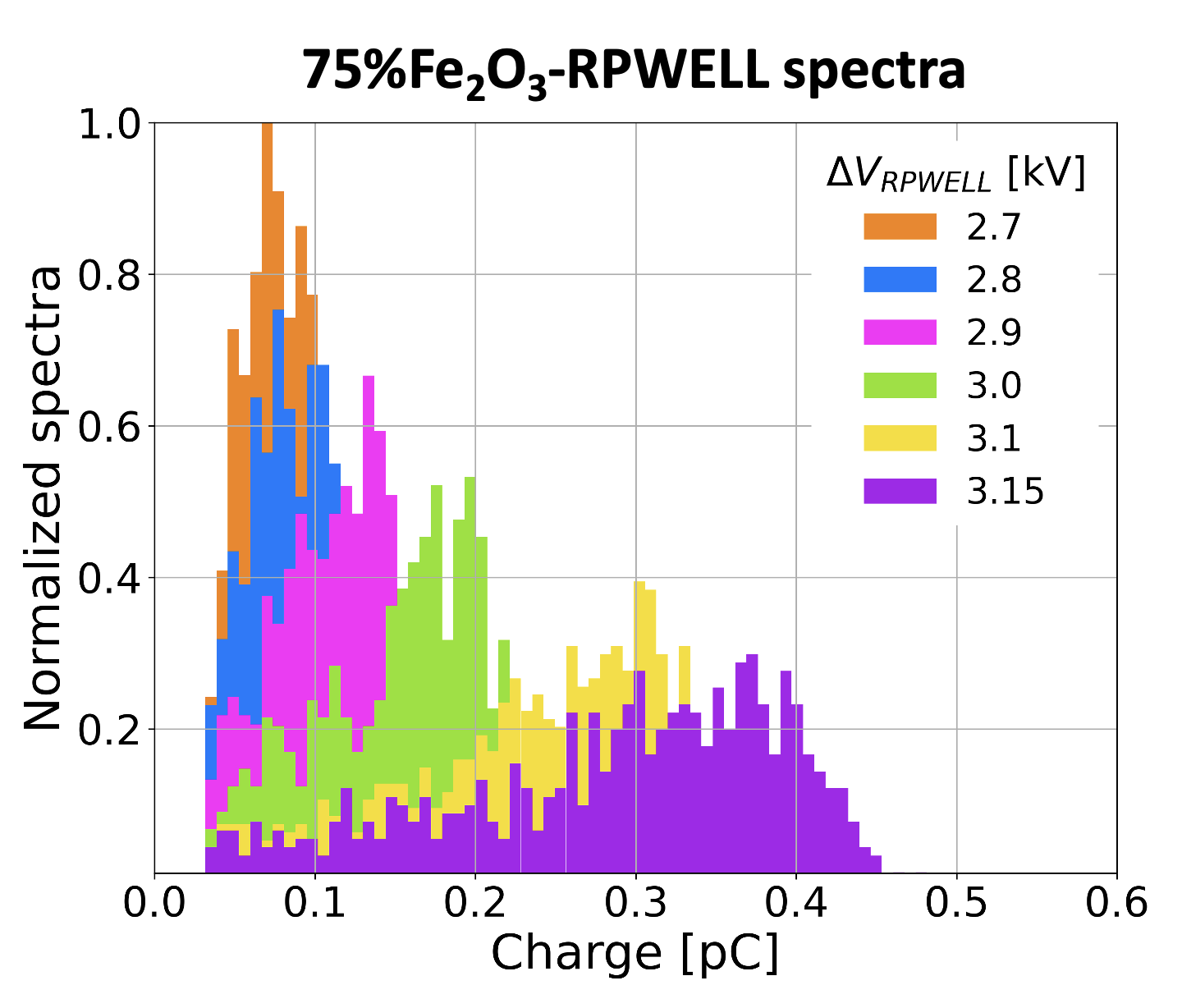}
    \caption{\footnotesize 
    Example of reconstructed $\alpha$-spectra obtained from a 75$\%$Fe$_2$O$_3$-RPWELL operated at 90K, 1.2 bar.}
    \label{fig:Amplification}
\end{figure}

In Fig.~\ref{fig:Amplification}, the normalized charge-amplitude spectra are shown. The calibration was performed by injecting a known amount of charge through a 2 pF capacitor. The mean value of each amplification spectrum was normalized over the mean value of the collection spectrum (see Fig. 5 of \cite{Tesi_2023_rwell}), the latter corresponding to $\sim$1.5 x 10$^5$ primary electrons.

The low-energy tail is attributed to the partial energy deposition of alpha particles exiting the collimator at large angles. 
With increasing voltage, the general shape of the spectra was preserved, with no indication of distortions due to secondary effects or discharges.\\

\noindent

\subsection{Stable gain curve and discharge probability}
\label{subsec:stable_gain}
\begin{figure}[h]
 \centering
    \includegraphics[width=7.5cm]{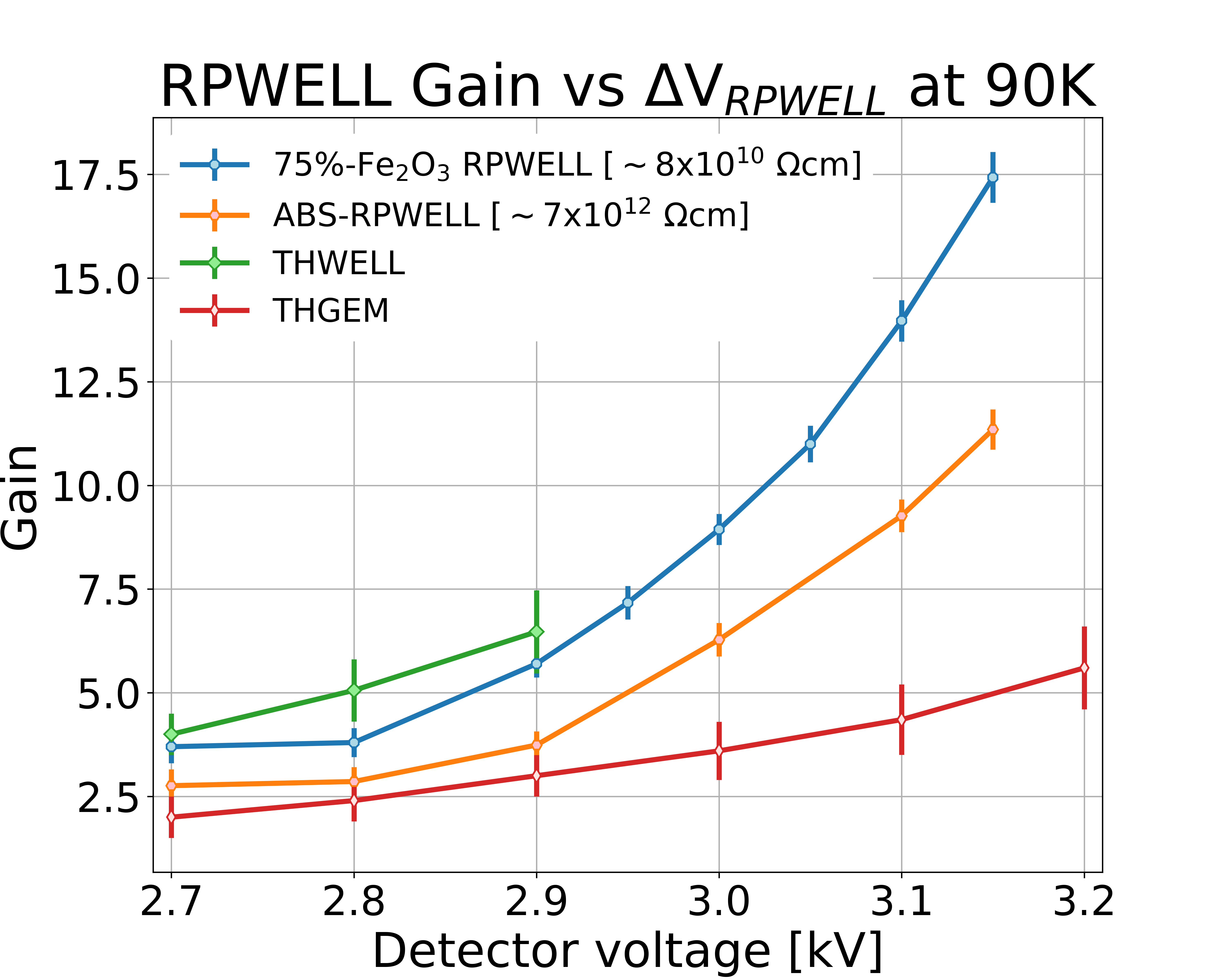}
    \includegraphics[width=7.5cm]{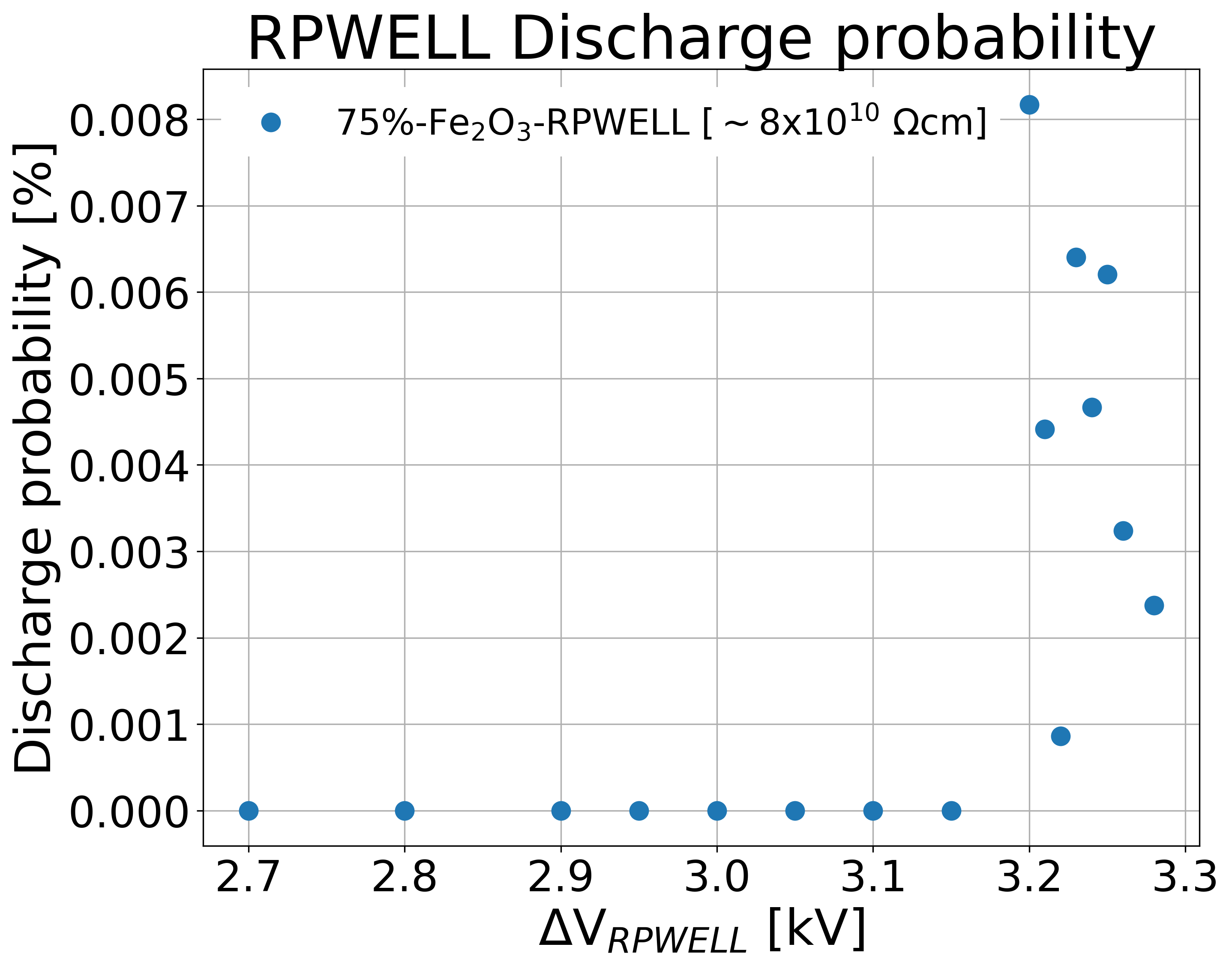}
    \caption{\textbf{Left}: stable gain as a function of $\Delta$ V$_\mathrm{RPWELL}$ for a 75$\%$ Fe$_2$O$_3$-RPWELL and ABS-RPWELL at 90~K, 1.2~bar in purified Ar. Data from THGEM and THWELL from \cite{Tesi_2023_rwell} are included for comparison; \textbf{Right}: discharge probability of the 75$\%$Fe$_2$O$_3$-RPWELL as a function of $\Delta$V$_{\mathrm{RPWELL}}$. }
    \label{fig:RPWELL_stableG}
\end{figure}

In what follows we consider only measurements carried out in stable gain conditions, i.e., after initial gain stabilization and prior to the appearance of discharges (Sec.~\ref{subsec:gain_stabilization}). 
Fig.~\ref{fig:RPWELL_stableG} left shows the gain as a function of $\Delta V_{\mathrm{RPWELL}}$ for a 75$\%$Fe$_2$O$_3$-RPWELL and for an ABS-RPWELL operated at 90~K, 1.2~bar of purified Ar. The results are compared to identical measurement carried out on non-resistive THGEM/THWELL \cite{Tesi_2023_rwell}. For all the investigated configurations, the gains exponentially increase with $\Delta V_{\mathrm{RPWELL}}$. As can be seen, the two resistive configurations outperform the non-resistive ones. A maximum gain value G~$\approx$17 was reached at $\Delta V_{\mathrm{RPWELL}}$ = 3.15~kV with the 75$\%$Fe$_2$O$_3$-RPWELL and G~$\approx$11 with the ABS-RPWELL at the same operational voltage, relative to a gain of 5 measured with the THGEM/THWELL in the same conditions.  
Beyond this value, the detector was still capable of operating in the presence of discharges, as described in the next section.\\

\noindent
The discharge probability is defined as the number of discharges over the total number of events. It was estimated by normalizing the number of discharges recorded from the power supply current monitor over the elapsed time and the source counting rate. The discharge probability as a function of $\Delta$V$\mathrm{_{RPWELL}}$ is depicted in Fig.~\ref{fig:RPWELL_stableG} right for a 75$\%$ Fe$_2$O$_3$-RPWELL at 90~K, 1.2 bar in purified Ar.

\noindent
One observes that in the stable region, of 2.7~kV < $\Delta$V$\mathrm{_{RPWELL}} \le$  3.15~kV, the detectors with the different plates are discharge-free (discharge probability < 0.005$\%$). A phase transition occurs around $\Delta$V$\mathrm{_{RPWELL}} \ge$ 3.15~kV. Beyond this voltage, a region of instability characterized by quenched discharges appeared. The latter occurred with a magnitude around 1-1.3 $\mu$C. The 75$\%$ Fe$_2$O$_3$-RPWELL could operate in the presence of quenched discharges albeit at an unstable gain condition. Beyond $\Delta V_{\mathrm{RPWELL}}$ = 3.28~kV, discharges led to a high DC current between the RPWELL-top and the grounded resistive plate, causing the tripping of the detector like in the non-protected ones. A similar behavior was recorded with the ABS-RPWELL. The described phenomenology could be caused by setup-related limitations, such as a bad coupling between the resistive plate and the THGEM due to non-perfect planarity. A similar behavior was recorded with the ABS-RPWELL. Additional investigation is ongoing to understand the origin of the aforementioned phenomenology.

\subsection{Summary and Discussion}
For the first time, a cryogenic RPWELL detector incorporating a YSZ+Fe$_2$O$_3$ ceramic plate (with Fe$_2$O$_3$ in the concentration of 75$\%$) was operated in argon vapor at cryogenic temperature (90~K, 1.2~bar). The detector was characterized with an alpha source. 
After an initial gain reduction of $\approx$60$\%$ imputable to charging-up, stable, discharge-free operation was recorded up to G~$\approx$17. Above the maximal stable gain, the detector was still operable but, due to the presence of discharges, the gain became unstable. It is interesting to remark that the discharges in the RPWELL had a magnitude $\approx$~3-fold smaller than the ones recorded for a THWELL ($\ge$~3.75~$\mathrm{\mu}$C) and they were able to destabilize the detector gain for $\approx$30~minutes. The transition from the discharge-free regime to the discharge one was sharp. 
Preliminary results for an ABS-RPWELL based on a CNT-doped 3D-printed thermoplastic plate were obtained reaching a stable charge gain of ~$\approx$11. These novel materials represent an alternative to ceramic: they have the advantage of having an almost temperature-independent bulk resistivity and are potentially easier to manufacture in large areas. However, it is known that thermoplastic materials may present problems of outgassing and aging that could hinder their adoption \cite{Chiggiato}. This matter requires further investigation.
In comparison to the results presented in \cite{Tesi_2023_rwell}, the cryogenic RPWELLs outperformed the non-protected configurations (THGEM, THWELL) of equal dimensions, operated in the same conditions. On the other hand, the cryogenic RWELL reached higher gains (up to $\approx$30) but at a cost of high discharge probability ($\approx$1$\%$). Both the RWELL and the RPWELL were found to have very low discharge probability, near zero, at equal gains of $\approx$17.\\

\noindent
The results presented in this work could motivate future studies in the direction of discharge-protected detectors based on resistive ceramic plates or CNT-doped ABS as a potential candidate for applications requiring charge multiplication in cryogenic conditions, e.g., in dual-phase liquid argon TPCs. 

\label{sec:conclusions}

\acknowledgments
\noindent
We would like to thank Dr. Yariv Pinto and Matan Divald from the 3D functional and printing center of the Hebrew University of Jerusalem for the production of the CNT-doped 3D-printed materials and Dr. Gregory Leitus from the Department of Chemical Research Support at the Weizmann Institute of Science for the assistance with the materials characterizations. This work was supported by Sir Charles Clore Prize, by the Nella and Leon Benoziyo Center for High Energy Physics, and by CERN-RD51 funds through its ‘common project’ initiative, and by the Pazy foundation. Special thanks go to Martin Kushner Schnur for supporting this research.
We acknowledge as well financial support from Xunta de Galicia (Centro singular de investigación de Galicia, accreditation 2019- 2022), and by the “María de Maeztu” Units of Excellence program MDM-2016-0692. DGD was supported by the Ramón y Cajal program (Spain) under contract number RYC-2015-18820. 
 
\section{Bibliography}
\bibliographystyle{JHEP}
\bibliography{bibliography}
\end{document}